\documentclass[11pt]{article}
\usepackage{moriond,epsfig}

\def\be{\begin{equation}}
\def\ee{\end{equation}}
\def\bea{\begin{eqnarray}}
\def\eea{\end{eqnarray}}

\def\fakelesssim{\:{{}^{{}_{\displaystyle <}}_{\displaystyle \sim}}\:}

\begin{document}
%%%%%%%% HAS BEEN SUPPRESSED \vspace*{4cm}
\title{COOPER PAIR BOX COUPLED TO A CURRENT-BIASED JOSEPHSON JUNCTION}

\author{\underline{F.W.J. HEKKING}$^1$, O. BUISSON$^2$, F.
BALESTRO$^{2}$ AND M.G. VERGNIORY$^1$}

\address{$^1$Laboratoire de Physique et Mod\'elisation des Milieux Condens\'es
\& Universit\'e Joseph Fourier, C.N.R.S., BP 166, 38042
Grenoble-cedex 9, France}
\address{$^2$Centre de Recherches sur les Tr\`es Basses Temp\'eratures,
laboratoire associ\'e \`a l'Universit\'e Joseph Fourier, C.N.R.S.,
BP 166, 38042 Grenoble-cedex 9, France}

\maketitle\abstracts{We study the dynamics of a quantum
superconducting circuit which consists of a Josephson charge
qubit, coupled capacitively to a current biased Josephson
junction. Under certain conditions, the eigenstates of the qubit
and the junction become entangled. We obtain the time evolution of
these states in the limit of weak coupling. Rabi oscillations
occur, as a result of the spontaneous emission and re-absorption
of a single oscillation quantum in the junction. We discuss a
possible way to experimentally determine the quantum state of the
junction and hence observe the Rabi oscillations.}

\section{Introduction}
\label{introduction}

Advances in quantum information theory~\cite{Steane98} are at the
origin of the recent search for quantum mechanical two-level
systems that can be used as quantum bits (qubits). In order to
process quantum information in a useful way, certain operations on
systems which consist of many qubits must be performed, such as
the preparation and manipulation of, as well as a measurement on,
entangled quantum states of coupled qubits. In this context,
superconducting devices using small Josephson junctions are of
particular interest. It has been experimentally demonstrated that
a single Cooper pair box constitutes a two-level system which can
be coherently controlled~\cite{Bouchiat98,Nakamura97,Nakamura99}.
Moreover, the use of a Cooper pair box as a qubit in the context
of quantum computers was studied
theoretically~\cite{Shnirman97,Makhlin99,Makhlin00}. But as of
yet, the existence of entangled states, which are at the heart of
quantum information processing, has not been demonstrated
experimentally.

In this article we study one of the simplest superconducting
circuits in which entangled states can be
realized~\cite{Buisson00}. It consists of a Cooper pair box coupled to
a current-biased Josephson junction. Theoretically, this
quantum circuit can be described by a two-level system coupled to
a harmonic oscillator. After a description of the quantum circuit
in the next Section, we discuss the time evolution of its
eigenstates in Sec.~\ref{entanglement}. In particular, we
demonstrate the existence of entanglement, very similar to the
entanglement found for an atom in an electromagnetic
cavity~\cite{Brune96}. In Sec.~\ref{measure} we discuss a possible
way to measure the Rabi oscillations, associated with the dynamics
of the entangled states. The last Section contains a discussion of
the feasibility of our proposal.

\section{Superconducting circuit}
\label{circuit}

\begin{figure}[thpb]
\centerline{\psfig{file=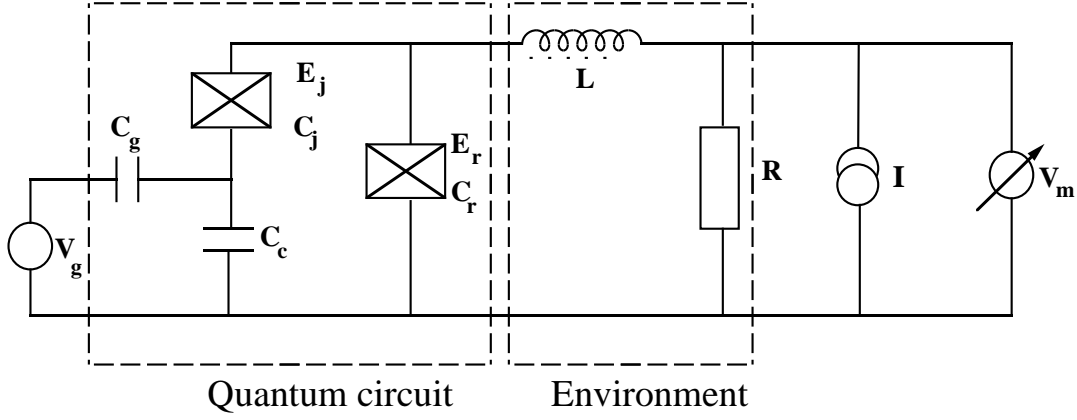,height=5.5cm}}
\caption{Superconducting quantum circuit connected to an
electrodynamic environment.} \label{device}
\end{figure}

The circuit that we will study is depicted in
Fig.~\ref{device}. It contains a quantum superconducting
circuit connected to an
electrodynamic environment. The quantum circuit itself
consists of three elements,
which we will discuss in some detail below.

\subsection{Cooper pair box}
The Cooper pair box consists of a small superconducting island,
coupled capacitively to a gate voltage $V_g$ (gate capacitance
$C_g$). The island is furthermore connected capacitively to a
superconducting electrode via a Josephson junction (capacitance
$C_j$, Josephson energy $E_j$). Using the basis states
$|n\rangle$, where $n$ corresponds to the number of excess Cooper
pairs on the island, we can write the Hamiltonian for the box as
\be \hat{H}_{\rm box} = E_{C,j} \sum \limits_{n} (2n - N_{g})^{2}
|n\rangle \langle n| - \frac{E_{j}}{2} \sum \limits_{n}
\left(|n+1\rangle \langle n| + |n-1 \rangle \langle n| \right).
\ee Here, $E_{C,j}=e^2/2C_{j,\rm eff}$ is the total charging
energy of the box ($C_{j,\rm eff} = C_j +[1/C_r +
1/(C_c+C_g)]^{-1}$) and $N_g = -C_{g}V_{g}/e$. The first term of
this Hamiltonian corresponds to the electrostatic energy of the
box. Generally, this energy is minimal if the number $n$ of excess
Cooper pairs on the island is an integer. However, if the
dimensionless gate-voltage $N_g$ is an odd integer, $N_g = 2m+1$,
a degeneracy occurs, and the number $n$ of Cooper pairs fluctuates
between $m$ and $m+1$. The second term of the Hamiltonian,
corresponding to the Josephson coupling between the island and the
electrode, lifts this degeneracy. As a result, close to odd
integer values of $N_g$, a gap opens up of order $E_j$ and the
relevant charge states are superpositions of the charge states
$|m\rangle$ and $|m+1\rangle$. In particular, if the gate-voltage
is such that $N_{g} \simeq 1$, the states with $n=0$ and $n=1$ are
almost degenerate. At low energies, the Hamiltonian $\hat{H}_{\rm
box}$ involves only these two states, and thus can be written as a
matrix
\begin{equation}
\hat{H}_{\rm box} \simeq \left(
\begin{array}{cc}
E_{C,j} N_{g}^{2} &-E_{j}/2 \\
-E_{j}/2   &E_{C,j} (2-N_{g})^{2}
\end{array}
\right).
\end{equation}
The two eigenstates $|-\rangle$ and $|+\rangle$ are superpositions
of the charge states $|0 \rangle$ and $|1\rangle$,
\begin{eqnarray}
|-\rangle &=& \alpha |0\rangle + \beta |1\rangle \\
|+\rangle &=& \beta |0\rangle - \alpha |1\rangle ,
\end{eqnarray}
where $\alpha ^{2} = 1- \beta ^{2} = [1+ \delta E_{g}/\sqrt{
(\delta E_{g})^{2}+ E_{j}^{2}}]/2$ with $\delta E_{g} = -4E_{C,j}
\delta N_{g}$ and $\delta N_{g} = N_{g} -1$. The corresponding
eigenenergies are
\begin{equation}
E_{\mp} = E_{C,j}[1+(\delta N_{g})^{2} ] \mp \frac{1}{2}
\sqrt{(\delta E_{g})^{2} + E_{j}^{2}}.
\end{equation}
We see that the Cooper pair box behaves as a quantum mechanical
two-level system,
usually referred to as a Josephson charge qubit

\subsection{Current biased Josephson junction}
A current biased Josephson junction can be characterized by two
conjugate variables: the charge $\hat{Q}$ on the junction and the
phase difference $\phi$ across it. Ignoring
effects related to the presence of the
environment, we can write the Hamiltonian as \be \hat{H}_r =
\frac{\hat{Q}^2}{2C_{r,\rm eff}}+ U(\phi) ,\ee where $C_{r, \rm
eff}$ is the effective capacitance of the junction in the circuit,
$C_{r,\rm eff} = C_r +[1/C_j + 1/(C_c+C_g)]^{-1}$, and $U(\phi) =
- E_r \cos \phi - ({\Phi_{0}/2\pi}) I \phi $ is the well-known tilted
washboard potential ($\Phi_{0} = h/2e$ is the superconducting flux
quantum). It depends on the Josephson energy $E_r$ and the bias
current $I$. For bias currents below the Josephson critical
current $I_c = 2\pi E_r/\Phi _0$, the potential $U(\phi)$ contains
local periodic minima as a function of $\phi$. These minima are
quadratic with characteristic frequency $\omega_r = (\sqrt{8E_r
E_{C,r} }/\hbar) (1-(I/I_c)^2)^{1/4}$, where $E_{C,r} = e^2/2C_{r,
\rm eff}$.

For very low bias currents ($I\ll I_{c}$), the minima are separated by
large potential barriers $\Delta U \sim 2E_r-\Phi_{0}I/2$. As a
result, in the quantum limit, various low-lying states can be
found in each minimum if $E_r \gg E_{C,r}$; the broadening of the
energies of these states due to tunneling between neighboring
minima can be ignored. Hence these states are well approximated by
harmonic oscillator eigenstates, {\em i.e.}, we can approximate $
\hat{H}_r \simeq \hbar \omega _r (a^\dagger a + 1/2)$, where the
operator $a^\dagger$ ($a$) creates (destroys) an oscillation
quantum in the junction. The eigenstates are denoted
$|\chi_k\rangle$, corresponding to the presence of $k= 0, 1, 2,
\ldots$ oscillation quanta in the junction. Thus, at low bias
current, the junction behaves as a quantum mechanical
superconducting resonator. Since in this limit the phase is
localized in a well-defined minimum of the potential $U(\phi)$,
the time-averaged voltage $V_m$ across the junction remains zero.

If the bias current is increased to values close to $I_c$ such
that $I \fakelesssim I_c$, the potential $U(\phi)$ is well
approximated by a cubic potential with a barrier height given by
$\Delta U \simeq (4\sqrt{2}/3) E_{r}(1-I/I_c)^{3/2}$. Both $\Delta
U$ and $\hbar \omega_r$ decrease with increasing $I$; they vanish
when $I=I_c$. The number of localized states in a given well
decreases as $\Delta U/ \hbar \omega_r
\sim(1-I/I_c)^{5/4}$. Moreover, the remaining levels
broaden due to quantum tunneling out of the minima of the potential
$U(\phi)$. A tunneling event changes the state of the
junction~\cite{Devoret85}, thereby leading to a finite voltage $V_m$
across it that can be detected.
The broadening of the ground state energy is given by
the tunneling rate $\Gamma _0 \sim \hbar \omega_r \exp(- \Delta
U/\hbar \omega_r)$. Since the excited states are located closer to
the top of the barrier, the tunneling rates increase with
increasing level index $k$: $\Gamma_k/\Gamma_ {0} \sim (432\Delta
U/ \hbar \omega_r)^k$~\cite{Weiss}. The tunneling rates for the
first two states $|\chi_0\rangle$ and
$|\chi_1\rangle$ are therefore very different. We will see in
Section~\ref{measure} how this property can be used in order to
distinguish these two quantum states.

\subsection{Coupling capacitance}
The capacitance $C_{c}$ plays a crucial role in the circuit of
Fig.~\ref{device} since it couples the charge qubit and the
current-biased junction to each other. As a result, these two
circuits are no longer independent and the system must be
considered in its totality. Physically, the capacitance $C_c$
couples the charge $2n-N_g$ on the qubit to the charge $Q$ on the
junction. Using the relation $\hat{Q} = \sqrt{\hbar \omega_r
C_{r,\rm eff}/2} (a - a^\dagger)/i $, the coupling Hamiltonian can
be written as
\begin{equation}
\hat{H}_{c} = - i E_{\rm coupl}  (2 n - N_{g}) (a^\dagger - a) .
\label{Hc}
\end{equation}
Here we introduced the characteristic coupling energy $E_{\rm
coupl} = \sqrt{\hbar \omega _{r}/E_{C,r}} E_{C,c}/2$, where
$E_{C,c} = e^2/2C_{c, \rm eff}$, with $C_{c, \rm eff} =
C_c+C_g+(1/C_j+1/C_r)^{-1}$. We will see in the next section how
this coupling energy leads to entanglement between the states of
the qubit and the junction.

\section{Operation at low bias current: entanglement}
\label{entanglement}

In this section we discuss the behaviour of the circuit for values
of the bias current away from the critical current, $I \ll I_c$.
In this limit, the system can be considered as a two-level system
(the qubit) coupled to a harmonic oscillator (the current biased
junction). The dynamics of such a system has been discussed in the
past~\cite{Jaynes63}. More recent applications are concerned with
quantum optics~\cite{Brune96} or Josephson junctions
\cite{Marquardt00}. Here we will show how entangled states can be
obtained which involve both the qubit and the
junction~\cite{Buisson00}.

Throughout this section we will be interested in the situation
$N_{g} \simeq 1$, such that we have to consider the qubit states
$|-\rangle$ and $|+\rangle$ only. Furthermore, as far as the
junction is concerned, we will limit the discussion to the ground
state $|\chi_0\rangle$ and the first excited state $|\chi_1
\rangle$. Hence, in the absence of coupling, $E_{\rm coupl} = 0$,
the four lowest energy eigenstates are $| -, \chi_0\rangle, | -,
\chi_1\rangle, | +, \chi_0\rangle$, and $| +, \chi_1\rangle$. Let
us now consider the special case $\hbar \omega _r = E_j =
\bar{E}$, {\em i.e.}, the states $| -, \chi_1\rangle$ and $| +,
\chi_0\rangle$ are degenerate for $N_{g} = 1$. Their energy is
$\bar{E}$ with respect to the energy of the ground state $| -,
\chi_0\rangle$. In the limit of weak coupling, $E_{\rm coupl} \ll
\bar{E}$, this degeneracy is lifted: the states $| -,
\chi_1\rangle$ and $| +, \chi_0\rangle$ get mixed and hence give
rise to the two entangled states $|\psi_1\rangle$ and
$|\psi_2\rangle$,
\begin{eqnarray}
|\psi_1\rangle &=& [|-,\chi_1\rangle +i
|+,\chi_0\rangle]/\sqrt{2},
\nonumber \\
|\psi_2\rangle &=& [|-,\chi_1\rangle -i
|+,\chi_0\rangle]/\sqrt{2}, \nonumber
\end{eqnarray}
corresponding to the energies $\bar{E} - E_{\rm coupl}$ and
$\bar{E} + E_{\rm coupl}$, respectively.

Suppose that the system has been prepared in the state $|\psi
(t=0)\rangle = |+,\chi_0\rangle$ at time $t=0$. This can be
achieved by a suitable manipulation of the gate voltage $V_g$ at
times prior to $t=0$~\cite{Nakamura99}. If we keep $V_g$ fixed
such that $N_g = 1$ at times $t>0$, the time evolution of $|\psi
(t) \rangle$ is given by
\begin{equation}
|\psi (t) \rangle = \frac{1}{\sqrt{2}i} \left[ e^{-i(\bar{E}
-E_{\rm coupl})t/\hbar} |\psi_1\rangle - e^{-i(\bar{E} + E_{\rm
coupl})t/\hbar} |\psi _2\rangle \right] .
\end{equation}
We see that the state $|\psi (t) \rangle$ oscillates coherently
between $|-,\chi_1\rangle$ and $|+,\chi_0\rangle$. In fact, these
so-called quantum Rabi oscillations can be interpreted as the
spontaneous emission and re-ab\-sorp\-tion of one oscillation
quantum by the junction. An interesting quantity is the
probability $P_{\chi_0} (t)$ to find the junction in the state
$|\chi_0\rangle$ (no oscillation quanta) after a certain time $t$.
This probability shows Rabi oscillations as a function of $t$ with
frequency $2E_{\rm coupl} /\hbar$,
\begin{equation}
P_{\chi_0} (t) = |\langle +,\chi_0|\psi(t)\rangle|^2 = \frac{1}{2}
[1 + \cos (2E_{\rm coupl} t/\hbar)] . \label{P1}
\end{equation}
Since these Rabi oscillations are characteristic for the
entanglement realized in the circuit, their measurement would
provide direct evidence of the presence of the entangled states
$|\psi_1\rangle$ and $|\psi_2\rangle$. We will discuss the
possibility to perform such a measurement in the next Section.

\section{Operation at bias current close to $I_c$: quantum measurement}
\label{measure}

In order to measure the probability $P_{\chi_0} (t)$ to find the
junction in the state $|\chi_0\rangle$ at time $t$, it is
necessary to perform a series of measurements. Each of these
measurements consists of the preparation of the state $|\psi
(t=0)\rangle = |+,\chi_0\rangle$ at time $t=0$ and the subsequent
coherent evolution until the time $t$, when the actual measurement
of the state of the junction is performed. The statistics of the
ensemble of measurements then yields $P_{\chi_0} (t)$.

The proposed measurement procedure consists in transforming the
two distinct quantum states $|\chi_0\rangle$ and $|\chi_1\rangle$
into two different stable classical states which are easily
measurable: the zero-voltage and the finite voltage state of the
current-biased junction. This measurement is a one shot method which gives binary
information on the initial quantum state of the junction.

In order to measure the state of the junction at a certain time
$t$, we propose to proceed as follows. Starting at time $t$, the
bias current $I$ is increased during a short ramping time $\delta
t$ to a value $I_{m}$ close to $I_c$. The bias current is kept at
$I_{m}$ for a time $\Delta t$ and then switched back to a low
value away from $I_{c}$. Finally, the DC-voltage across the
junction is measured.

The increase of the bias current to $I_{m}$ drastically increases
the tunneling rates $\Gamma_{0}$ and $\Gamma_{1}$ of the two
possible states of the junction. During the time $\Delta t$, the
junction can transit into a finite voltage state as a result of a
tunneling process. However, as we have seen, the escape time
depends strongly on the initial quantum state. For instance,
$\Gamma _1 \sim 1000 \Gamma _0 $ if $\Delta U/ \hbar \omega_r > 3$.
Therefore, if the junction is in the state $|\chi_1\rangle$, a DC
voltage starts to develop after a time $1/\Gamma _1$, whereas this
will happen after a much longer time $1/\Gamma _0 \gg 1/\Gamma _1$
if the junction is in the state $|\chi_0\rangle$. Hence if we switch the bias
current $I$ back to a value away from $I_c$ after a time $\Delta
t$ such that $1/\Gamma _1 \ll \Delta t \ll 1/\Gamma _0$, the
junction remains in the zero-voltage state {\em if and only if the
junction is in the state} $|\chi_0\rangle$. It is during the time
$\Delta t$ that the two quantum states bifurcate into two
different voltage states, hence we refer to this time as a
measuring time. After $\Delta t$, the current is switched back to
a lower value. Since the $I-V$ characteristics of the junction are
hysteretic, the zero voltage and finite voltage states remain
dynamically stable. Thus we have a relatively long time to perform the actual
voltage measurements and  distinguish the two junction
states.

\section{Discussion}
\label{discussion}

The functioning of the quantum circuit presented in this paper
relies on various assumptions. Below we will discuss these
assumptions for each part of the circuit in some detail.

{\em Josephson charge qubit.} One of the main assumptions
concerning the qubit is the absence of quasiparticles. If a
quasiparticle tunnels onto the island, the state of the Cooper
pair box changes: it no longer behaves as a quantum mechanical
two-level system. Thus we need the
superconducting gap to be sufficiently large, $\Delta
> E_{C,j} + E_j/2 $. We furthermore need the box to be quantum mechanically
coherent during the experiment. This is one more reason to impose
the absence of quasiparticles. But in addition it means that the
coupling to the outside world must be sufficiently weak. Generally, the main
source of decoherence is formed by time-dependent fluctuations of the
effective gate charge $N_g$, either induced by fluctuations of the
gate-voltage $V_g$, or by residual dynamics of charged impurities
close to the qubit. However, if the qubit is operated close to the
degeneracy point $N_g = 1$, the effect of this noise will be weak,
since the energies $E_\pm$ depend only quadratically on the
amplitude of the fluctuations.

{\em Current biased Josephson junction.} When using the current
biased Josephson junction as a quantum mechanical resonator, it is
essential for the states $|\chi_0\rangle$ and $|\chi_1\rangle$ to be
well-defined,{ \em i.e.}, the broadening $\Gamma_{0,1}$ of the
corresponding energy levels must be small compared to $E_{\rm
coupl}$. For this reason we need to work at low bias current, such
that $\Gamma_1 \sim 1000 \Gamma_{0}\sim 1000 \sqrt{E_r E_{C,r}}
\exp{\sqrt{E_r/E_{C,r}}} \ll E_{\rm coupl}$. In addition,
decoherence to the external circuit should not occur on a
time scale of the order of $\hbar /E_{\rm coupl}$.  This means that
the impedance should be sufficiently high. For a purely resistive
environment for instance ($R$ finite and $L=0$ in Fig.~\ref{device}),
the typical relaxation time is $RC_r$.
We do require this time to exceed the Rabi
period $\hbar /E_{\rm coupl}$.

When using the junction as a detector it is important that it
changes its state before relaxation processes induced by
the environment occur. Thus we need to increase the current $I$
sufficiently close to $I_c$ such that $1/\Gamma_1$ is much smaller
than the relaxation time.

Apart from increasing the tunneling rates $\Gamma_k$, an increase
of the bias current $I$ has another important effect. Since the
frequency $\omega_r$ depends on $I$, the resonance condition
$\hbar \omega_r = E_j$ is no longer satisfied after the increase.
As soon as the
difference $|\hbar \omega_r - E_j|$ exceeds $E_{\rm coupl}$, the
qubit and the junction are no longer entangled and the states $|
-, \chi_1\rangle$ and $| +, \chi_0\rangle$ become again
eigenstates of the system. The probability to be in either of
these states $|\chi_0\rangle$ or $|\chi_1\rangle$ ceases to evolve
in time. This way, the probability $P_{\chi_0}$ remains "frozen"
into its value at the time $t$ of the measurement
(as long as relaxation phenomena can be
neglected). In order for $t$ to be well-defined, the ramping time
$\delta t$ must be short compared to the period $\hbar/E_{\rm
coupl}$ of the Rabi oscillations. On the other hand, the ramping
itself should not induce transitions in the junction, thus we need
to impose the condition $\omega_r \delta t \gg 1$.

In most experiments~\cite{Devoret85,Silvestrini97},
the environment is dominated by a resistor
$R$. On the one hand,
$R$ should be large in order for the relaxation time to be
long. On the other hand, $R$ should be low enough to avoid heating
at low temperature due to the bias current flowing through the
resistor. In order to
avoid these confliciting conditions, we propose to use a large
on-chip inductor in series with the junction (see
Fig.~\ref{device}). To minimize the effect of the environment, the inductor
has to be much larger than $R/\omega_{r}$ and $1/(C\omega_{r}^2)$.
Then the relaxation time is given by $(L\omega_{r})^2 C/R$, which can
be much larger than typical $RC_r$-times~\cite{Devoret85}; the
characteristic frequency of the junction is not affected by the
environment. Moreover, if the temperature $T$ of the environment satisfies the
condition $R/L\ll 2kT/\hbar \ll\omega_{r}$, the escape rate is given by
$\ln(\Gamma_{0}(T)/\Gamma_{0}) =(9/ \pi \sqrt{2}) (h/e^3)
(C_{r}/LI_{c}) (1-I_{m}/I_{c})^{1/2}kT$,
where $\Gamma_{0}$ is the
zero-temperature rate for quantum tunneling. The excess escape rate
due to thermal fluctuations in the resistor is, surprisingly,
independent of the resistance $R$ of the environment.

For the numerical estimates presented below we will consider
parameters of typical
superconducting circuits~\cite{Devoret85,Nakamura99}.
For the Josephson charge qubit we
have chosen $E_j = 26.1 \mu$eV and
$E_{C,j} = 70 \mu$eV.
The coupling capacitance
is chosen to be of the order of $C_j$, $C_c = 0.5$fF, yielding
$E_c = 256$neV. For a Josephson junction with a
capacitance of 6.35pF and a
critical current of $10\mu$A, the characteristic frequency is about
13GHz at zero bias current.

The resonant condition $\hbar \omega_r = E_j$ for the
charge qubit and the Josephson resonator is obtained
at a bias current of $9.4\mu$A.
The period of the Rabi oscillations is given by $T_\mathrm{Rabi} \approx $8 ns.
Taking into account the environment
of Fig.~\ref{device} with $R=8\Omega$ and L=10nH,
the relaxation time is about 136ns and thus
longer than the period of the Rabi oscillations.

During the measurement process, the bias current
is chosen to be $I_{m}=9.91\mu$A, corresponding to a
characteristic frequency of 4GHz. The relaxation
time is 51ns. For $T=20$mK, we estimate the escape time to
the voltage state to be $6\mu$s from the state $|\chi_0\rangle$ and
$7ns$ from the state $|\chi_1\rangle$. Taking $\Delta t=50ns$, we expect
the junction to bifurcate into the voltage state with a
probability larger than 99.9\% for the state $|\chi_1\rangle$
and smaller than 0.1\% for the state $|\chi_0\rangle$.

In conclusion, we proposed an efficient procedure
for single-shot quantum measurements of entangled states in a Cooper pair
box coupled to
a current-biased Josephson junction.

\section*{Acknowledgments}
We thank G. Falci, R. Fazio, and E. Paladino for interesting
discussions.

\end{document}